\documentclass[%
 reprint,
 amsmath,amssymb,
 aps,
]{revtex4-2}

\usepackage{graphicx}
\usepackage{dcolumn}
\usepackage{bm}
\usepackage{color}
\usepackage{xcolor}

\newcommand{\dHybridR}{{\it dHybridR}}
\newcommand{\orcid}[1]{\hspace{1mm}\href{https://orcid.org/#1}{\includegraphics[height=0.3cm,keepaspectratio]{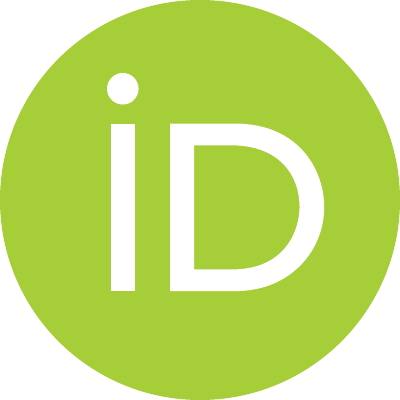}}}

\newcommand{\aap}{A\&A}

\newcommand{\mnras}{MNRAS}

\newcommand{\physrep}{Phys. Rep.}

\newcommand{\apss}{ApSS}

\newcommand\apjl{{ ApJL}}%

\begin{document}

\preprint{APS/123-QED}

\title{The role of non-linear Landau damping for cosmic-ray transport}

\author{Benedikt Schroer\orcid{0000-0002-4273-9896}}
\email{bschroer@uchicago.edu}
\affiliation{Department of Astronomy \& Astrophysics, University of Chicago, 5640 S Ellis Ave, Chicago, IL 60637, USA}

\author{Damiano Caprioli\orcid{0000-0003-0939-8775}} 
\affiliation{Department of Astronomy \& Astrophysics, University of Chicago, 5640 S Ellis Ave, Chicago, IL 60637, USA}
\affiliation{Enrico Fermi Institute, The University of Chicago, Chicago, IL 60637, USA}

\author{Pasquale Blasi\orcid{0000-0003-2480-599X}}
\affiliation{Gran Sasso Science Institute (GSSI), Viale Francesco Crispi 7, 67100 L'Aquila, Italy}
\affiliation{INFN-Laboratori Nazionali del Gran Sasso (LNGS),  via G. Acitelli 22, 67100 Assergi (AQ), Italy}

\date{\today}

\begin{abstract}
We present the first assessment, using hybrid PIC simulations, of the role of non-linear Landau damping in the process of self-generated scattering in a high $\beta$ plasma, conditions appropriate for CR scattering in the halo of the Galaxy. This damping process manifests itself in the form of heating of the background plasma and reduction of the drift speed of CRs that yet remains super-Alfvenic. We also show that the damping leads to an inverse cascade process, consisting of producing non-resonant large scale modes, a novel result with many potential phenomenological implications. 

\end{abstract}

\maketitle

\section{Introduction}
\label{sec:intro}
Modeling cosmic-ray (CR) transport in different environments requires a fundamental understanding of the interplay of various physical processes.
Based on measurements of secondary-to-primary CR flux ratios \cite{ams21,dampe22,CALET22}, CRs seem to diffuse due to scattering off turbulent magnetic fields.
Whether these fields are self-generated or are injected by supernova explosions on large scales and cascade from large to small scales is unclear. 
The discovery of a break in secondary-to-primary ratios \cite{ams21,dampe22,CALET22} might hint at CRs generating these fields via the excitation of streaming instabilities \cite{blasi+12,evoli+18b,chernyshov+24}, while pre-existing turbulence may take over at energies $\gtrsim\,$TeV.

The problem of CR transport in the presence of streaming instabilities excited by the same CRs is intrinsically non-linear as the growth rate depends on the CR distribution, which in turn is determined by the way particles propagate. 
Especially within the context of Galactic transport on kpc scales, the resonant streaming instability seems a promising candidate to describe CR confinement, at least up to $\lesssim\,$TeV energies \cite{blasi+12,evoli+18b} and perhaps even at all energies \cite{Dogiel2022}.

A crucial role in the assessment of the role of self-generation for CR transport is played by damping processes, which are thought to be responsible for saturating the growth of the instability and eventually determining the energy dependence of the diffusion process.

If growth occurs fast enough compared with damping, then the CR population has time to isotropize in the reference frame of the waves. Hence, their bulk speed $v_D$, initially super-Alfvenic to excite the instability, is reduced to the Alfvén speed $v_A$. 
While this may be important for low energy CRs, around GeV energies, the decreasing energy dependence of secondary-to-primary ratios suggests that higher energy particles move with a super-Alfvénic drift speed, $v_D>v_A$, that increases with energy. This is in fact expected if the growth of the resonant streaming instability is saturated due to damping processes \cite{kulsrud+71}. The common expectation is that waves of wavenumber $k$ grow due to the super-Alfvénic drift but that the level of perturbations is determined by the equality of the growth rate and the damping at the same wavenumber. Depending on the phase of the interstellar medium, the most relevant mechanisms are typically ion-neutral damping \cite{kulsrud+71} and non-linear Landau damping (NLLD) \cite{lee+73}, although many other processes might be at work \cite{hopkins+21,hopkins+22b, lemmerz+24, Cerri2024}.
In the hot ionized medium, the absence of neutrals renders NLLD the dominant process to limit the wave growth both on galactic scales \cite{lee+73} and even near sources \cite{nava+16,nava+19,dangelo+16}. 

The process can be pictured as follows: two Alfvén waves of slightly different wavenumber combine into a beat wave that can resonantly interact with particles of the background plasma, transferring energy from the waves to the plasma in the form of heat \cite{lee+73}. For linearly polarized waves, both waves are damped, and the background plasma is heated. For circularly polarized waves, the shorter wavelength wave transfers energy to both the other wave and the background plasma. The former case is typically considered more relevant because the resonant streaming instability of an almost isotropic CR distribution generates forward-moving left- and right-handed circularly polarized waves in nearly equal amounts. As a result, the waves are almost linearly polarized \cite{kulsrud+69}.

As discussed above, having in mind CR transport in our Galaxy, the streaming instability is assumed to be balanced by NLLD. In particular, the phenomenon of non-linear transport was studied on Galactic scales \cite{blasi+12,evoli+18b,chernyshov+24}, around SNRs \cite{lagage+83b,nava+19,dangelo+16}, in the multiphase ISM during Galaxy formation \cite{thomas+19}, in Galaxy clusters \cite{wiener+18}, in heating the Galactic gas \cite{wiener+13} and in launching Galactic winds \cite{Recchia2016,thomas+23,armillotta+24}.
On the other hand, numerical attempts at simulating the saturation of the instability focused mainly on the $\beta\ll 1$ regime in which NLLD is negligible or on ion-neutral damping, employing particle-in-cell (PIC) \cite{holcomb+19} or magnetohydrodynamics (MHD)+PIC \cite{bai+19,plotnikov+21} methods.

In this letter, we focus on the saturation of the resonant streaming instability in the astrophysically relevant regime of $\beta \gtrsim 1$ utilizing hybrid PIC simulations. 
We illustrate the first conclusive evidence that in such plasma NLLD is active and that it leads to a reduction of the drift speed of CRs driving streaming instability. Furthermore, contrary to previous studies we keep driving the instability instead of having a finite energy available in CRs with periodic boundaries.
This situation is reminiscent of a halo plasma element adjacent to the Galactic disk, that is constantly bombarded with freshly accelerated particles entering the plasma element from the disc region.

The article is structured as follows: in section \ref{sec:setup}, we describe the code and our simulation setup. Our results are presented in section~\ref{sec:results} and we conclude in section~\ref{sec:conclusions}.

\section{PIC simulations}
\label{sec:setup}
We study the self-confinement of CRs resulting from excitation of a resonant streaming instability and NLLD limiting the growth, using \dHybridR{}, a relativistic hybrid code with kinetic ions and (massless, charge-neutralizing) fluid electrons \citep{gargate+07,haggerty+19a}. We consider an adiabatic closure for the electron pressure, i.e., $P_e\propto \rho^{5/3}$.
\dHybridR{} is suited to properly simulate CR streaming instabilities \citep{haggerty+19a,schroer+21,schroer+22,zacharegkas+24}
and should capture all the relevant mechanisms for saturation of the resonant streaming instability, including NLLD.
By choosing a hybrid code, we are able to simulate longer time scales compared to PIC codes, essential for exploring low growth rates of the resonant streaming instability.

In our simulations, velocities, time and length scales are normalized to the Alfvén speed $v_A$, the ion cyclotron time $\Omega_0$ and the ion skin depth $d_i= v_A/\Omega_{0}$ respectively. Magnetic fields are normalized to the background magnetic field $B_0$. Particle number densities are normalized to the initial density of the background plasma, $n_0$.

Our simulation box is quasi-1D along $x$ and retains all three components of the momenta and electromagnetic fields. The total size is $60000\,d_i \times 5\,d_i$, divided into $120000 \times 10$ cells. The speed of light is fixed to $c=100\,v_A$ and the background magnetic field is directed along the $x$-axis with strength $B_0$. The background plasma is periodic in all directions and is sampled from a Maxwellian with an ion plasma $\beta=2 v_{th,i}^2/v_A^2=2$ and $128$ particles per cell, in order to reduce the magnetic field noise floor.
The electromagnetic fields are allowed to leave the simulation box, having open boundaries along the $x-$direction in order to prevent the recycling of the same waves. CRs have periodic conditions in $y$, open boundaries in $x$ and we use $16$ particles per cell.

The injected CR distribution is isotropic in a CR rest frame with $p_{iso}=100\,m v_A$ (Lorentz factor $\sqrt{2}$) and boosted into the simulation frame with $p_{bst}=5\,mv_A$ leading to a drift speed in the background plasma of $4\,v_A$, i.e., mildly super-alfvénic.
The time step is $0.005\,\Omega_{ci}^{-1}$.
Though some of the chosen parameters are not strictly equal to the values in the Galaxy (e.g., $c/v_A$), our setup respects the hierarchy of physical scales, which allows us to properly study the problem.

Our simulation box is $600$ CR gyroradii large, $r_L=100\,d_i$, to accurately resolve the maximally growing wavenumber $k_{res}=r_L^{-1}$ and to capture the scattering mean free path for magnetic fields as low as $\delta B^2/B_0^2\approx 2\times 10^{-3}$.
The instability is expected to grow with a growth rate $\gamma_{max}=\frac{\pi}{4}n_{CR}(v_D-v_A)\approx 4.4\times 10^{-4}\,\Omega_0$ for our distribution. 
An important detail in our setup is the implementation of the injector for the CR distribution. Due to CRs having $v_{iso}\gg v_D$, we have to inject particles at both open boundaries. 
In this way, we compensate for particles that would stream through the right-hand-side boundary into the box if the box were bigger than $L$.
However, with injectors fixed in time, once $v_D$ in the box decreases due to scattering, less particles are streaming back from the right-side than what would be expected for a larger box. We tested what happens when adjusting the injector to the effective drift speed in the box and concluded that our boundary conditions do not affect the main conclusions presented here, at least qualitatively.

\section{Results}
\label{sec:results}
\subsection{Evolution of the Magnetic Field}

\begin{figure}
    \centering
    \includegraphics[scale=0.5]{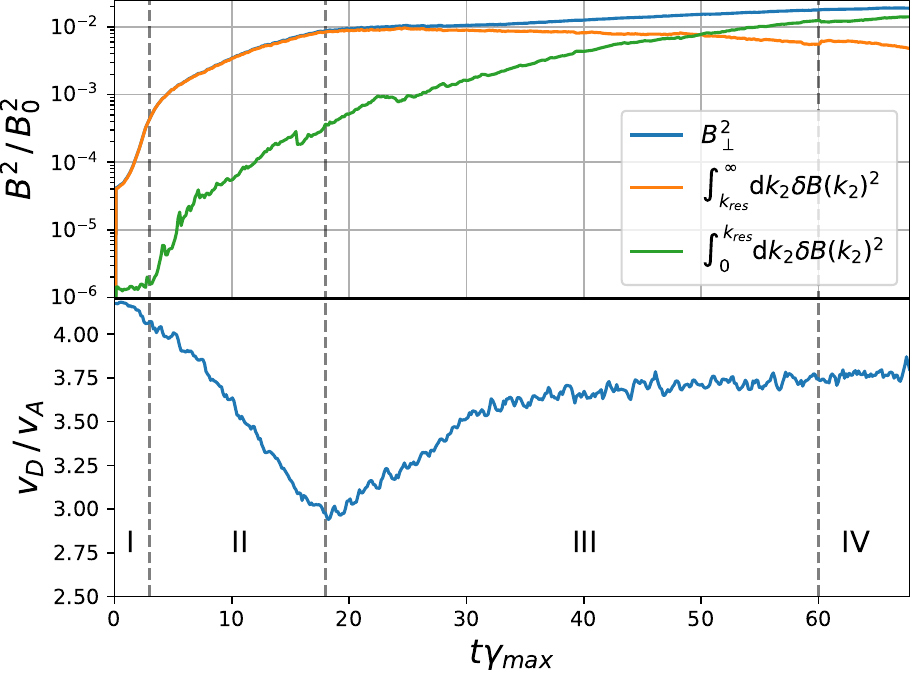}
    \caption{Bottom panel: Box averaged CR drift velocity as a function of time. Top panel:  Magnetic field on all scales smaller than $r_L$ (orange), larger than $r_L$ (green) and total perpendicular (blue) as a function of time. Time is in units of $\gamma_{max}$. Dashed lines indicate the transition of different evolutionary stages.}
    \label{fig:Bres}
\end{figure}

The overall evolution of the CRs and the magnetic field is illustrated in Fig.~\ref{fig:Bres}. We can identify four phases, that indicate different interplay of CRs and fields. In Stage I, CRs stream freely through the box with the initialized drift speed. The magnetic field's growth rate agrees well, within $10\,\%$, with the prediction derived from \cite{kulsrud+69}.
Around $t=3\,\gamma_{max}^{-1}$, the drift speed starts to drop as a result of CRs scattering off the self-generated magnetic perturbations (Stage II). The transition occurs when the mean free path is comparable to our box size, i.e., $\delta B^2/B_0^2\approx 10^{-3}$, see eq.~\ref{eq:mfp}.

At this point, scattering influences the evolution of $B^2_\perp$ by reducing the growth rate from the unperturbed one of our initial, pristine CR distribution to a new one given by the self-scattered distribution.

After $\sim 17\gamma_{max}^{-1}$, Stage III starts in which the field growth slows down significantly. In the absence of damping, the fields would keep growing and scattering would increase, until CRs become isotropic in the reference frame of the waves and get advected with $v_D=v_A$.
Hence, the superalfvénic drift throughout this stage with $v_D\gtrsim 3\,v_A$, indicates that other processes are opposing the growth of the waves. This point is even more evident when looking at the top panel of Fig.~\ref{fig:Bres}.

In Stage III, the magnetic field contained on all scales that are able to resonantly scatter CRs plateaus and then slightly damps away, despite $v_D>v_A$. After $ t\approx 60\,\gamma_{max}^{-1}$, the total field and the drift speed reach a plateau indicating the final phase of saturation (Stage IV).

\subsection{Non-Linear Landau Damping}
\begin{figure}
    \centering
    \includegraphics[scale=0.5]{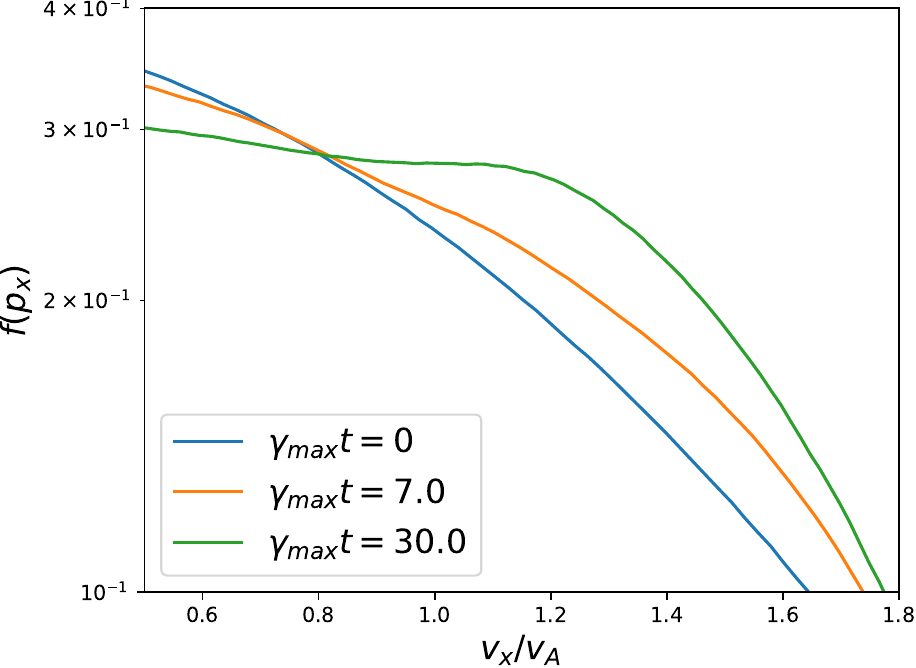}
    \caption{Background ion distribution function as a function of $p_x$ at three different times, showing a flattening at $p_x=m\,v_A$ as a signature of NLLD.}
    \label{fig:flattening}
\end{figure}

To understand the increasing drift speed observed in Stage III, it is instructive to first determine the reason why the power on resonant scales saturates. Fig.~\ref{fig:flattening} shows the background ion distribution function at three different times. At momenta around $p_x=m\,v_A$, the whole distribution flattens which is a clear signature of NLLD. As illustrated, the feature is already present around $t\approx 7\,\gamma_{max}^{-1}$, meaning that damping started to transfer energy from waves to the plasma before that, but remains too weak to compensate the growth due to CR streaming during Stage II. However, as the field continues to grow, damping can eventually saturate the field at different scales, setting the beginning of Stage III.

Naively we would expect fields to grow only on scales that can be resonant with the driving particles.
This expectation holds true initially, as can be seen from the power spectra of right- and left-handed waves in Fig.~\ref{fig:power_spectrum} and the large-scale fields being flat in Fig.~\ref{fig:Bres}.
However, once NLLD starts to operate, it is evident that the field starts growing on scales that are not in resonance with any of the particles. 
Furthermore, contrary to common approximations \cite{wiener+13,recchia+16,recchia+22,nava+19}, we confirm that small scales are damped by all the power contained on larger scales. This renders the damping rate maximal at small scales and gradually decreasing with $k$, compatible with earlier predictions \cite{volk+82};
more precisely, the power at small scales saturates first and starts partially damping away during Stage III (see Fig.~\ref{fig:power_spectrum}).
One subtle point is that the magnetic field on large scales keeps growing throughout Stage III. Hence, it is far from trivial that the damping rate exactly compensates the growth, and the resonant field is not entirely damped away.
Since the growth rate in our simulation changes due to scattering, it is impossible to determine the exact functional form of the damping rate and whether non-linear effects, such as particle trapping \cite{volk+82}, modify the predictions of linear analyses. Despite these issues, that deserve further investigation, there is enough evidence that NLLD is at work in our simulation and that it plays an important role to limit the growth of the waves. Overall, it is clear that magnetic power is transferred from small scales to large scales.

Interestingly, this inverse cascade is a feature of NLLD that was discussed in the early literature on this topic \citep{lee+73}, although it only occurs when two circularly polarized waves of the same polarization interact and to our knowledge has never been shown in hybrid simulations of the resonant streaming instability.
Typically, the resonant streaming instability creates linearly polarized waves, so naively this effect should be absent.
However, there is a small anisotropy of left- to right-handed modes of the order of $v_D/c$, since there are more particles with positive pitch angle than with negative one.
In Fig.~\ref{fig:power_spectrum}, one can see that initially the two helicities grow almost identical, i.e., the field is linearly polarized.
However, the small anisotropy is sufficient to halt the growth of the right-handed modes and render the final spectrum partially circularly polarized.
This kickstarts an inverse cascade that becomes efficient enough to remove power from small scales and relocate it to scales larger than the CR gyroradius.
As a result, opposite to what is always assumed, the energy given by CRs to the fields is not entirely dumped into heating the background gas, but rather it is partially transferred to large-scale modes for which damping is inefficient.
In terms of CR scattering, this opens the possibility for low energy CRs to influence the transport properties of higher-energy particles, with potentially profound implications for our description of CR transport.
We defer to a future paper the study of this effect.

\subsection{Drift Velocity}
\label{sec:drift}
\begin{figure}
    \centering
    \includegraphics[scale=0.5]{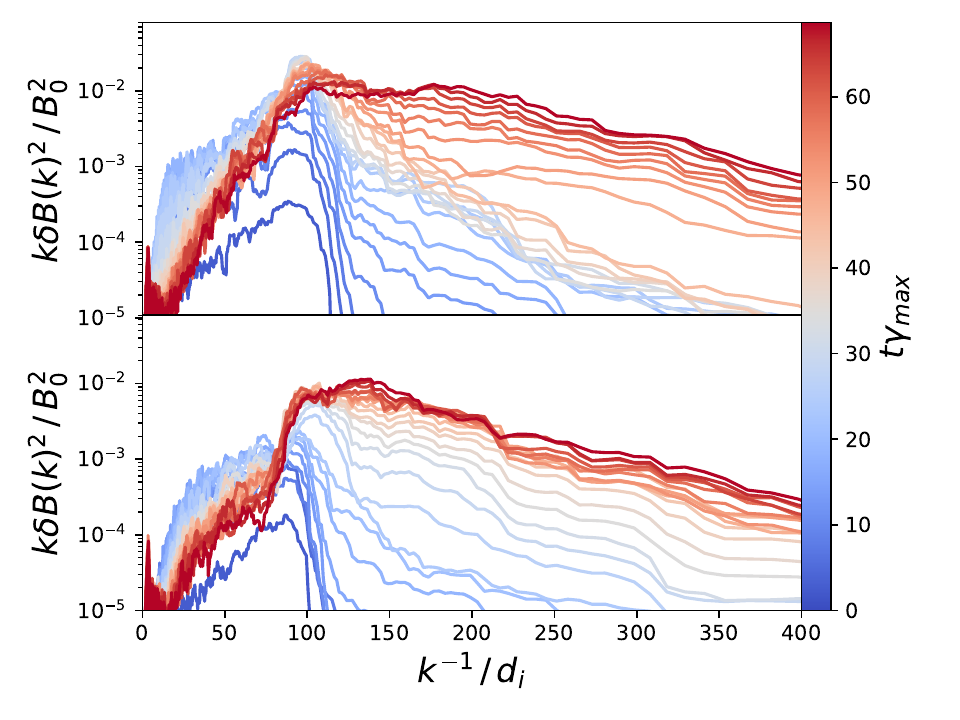}
    \caption{Power spectra of the magnetic field $\delta B_y \mp \mathrm{i}\delta B_z$at different k for left-handed (top panel) and right-handed (bottom panel) modes, respectively. Small-scale power is transferred to large scales, indicating an inverse cascade.}
    \label{fig:power_spectrum}
\end{figure}

The removal of power from small scales is reflected in the drift speed because it effectively worsens the so-called $90^\circ$-scattering problem, meaning that particles cannot cross the pitch angle barrier at $\mu=0$ and do not manage to isotropize. Overall, the increase in $v_D$ can then be understood as the pitch angle barrier becoming increasingly difficult to cross due to damping removing power from small scales. We can estimate the mean free path of a particle with a given $\mu=\frac{1}{k r_L}$ by calculating the product of its velocity along $x$ and the quasi-linear theory scattering time resulting in 
\begin{equation}\label{eq:mfp}
    \lambda = \frac{\mu c}{\Omega} \frac{2 B_0^2}{ \pi k\delta B(k)^2}\,.
\end{equation}
The particles efficiently scatter if $\lambda$ is smaller than the box size $L$ which means we need a magnetic field at the scale resonant with a given $\mu$ of $\sim 10^{-3}$.
From Fig.~\ref{fig:power_spectrum}, it is clear that during Stage II the field at small scales grows sufficiently large to ensure pitch angle scattering over a large range of values of $\mu$. However, at late times it gets damped away below this threshold which is why the drift speed starts to increase, as the pitch angle barrier becomes effectively too large to cross.

After the removal of power on small scales, the increase in drift speed slows down and approaches a new equilibrium value $v_D\sim 3.7\,v_A$, given by diffusion in the almost constant magnetic fields at the resonant scales. However, the power in large-scale fields is unexpectedly still increasing during that stage and even surpasses the power on resonant scales around $t\sim 50\,\gamma_{max}^{-1}$.

Since waves are allowed to leave the box, we expect that an Alfvén wave leaves the box at most after a time $L/v_A\sim 26\,\gamma_{max}^{-1}$. On the other hand, the energy transfer of CRs to magnetic fields and subsequent redistribution as heat and large-scale fields is a highly non-linear phenomenon depending on the fields at resonant scales and the CR drift speed among other parameters. As such, saturation in our box can only be achieved with a delay with respect to all these other processes, a delay that can be estimated as roughly one crossing time. Thus, we expect the green line to flatten with a delay of $26 \,\gamma_{max}^{-1}$ with respect to the drift speed and the peak of the orange line which would be around $t\sim 56\, \gamma_{max}^{-1}$. 
Indeed, the growth halts around $t\sim 60 \,\gamma_{max}^{-1}$ initiating Stage IV and a steady state is reached throughout our box.

At this point, the inverse cascade ends due to the waves leaving the box. In a closed periodic box, where waves cannot leave the box and are recycled on the other side, saturation is impossible to achieve, as we will discuss in detail in future work.

The configuration in Stage IV is qualitatively what we would expect to happen in the Galaxy: A CR distribution that is shaped by diffusion due to self-generated turbulent magnetic fields. 

A crucial new ingredient is the appearance of the large-scale fields, a by-product of self-generation. At present, it is not clear what could be the fate of such large scale modes: in principle, these waves could be {\it absorbed} resonantly by higher energy particles, that in this way may limit their growth. The spread in momentum in our simulation is too small to address this issue. It is however a point of major interest since in models of Galactic CR transport, large-scale turbulence is typically assumed to be injected only by supernova explosions \cite{blasi+12,evoli+18b} or produced by high-energy CRs in resonance with these scales \cite{chernyshov+24}.

Instead, our simulations indicate that large-scale fields naturally arise from NLLD even in the absence of these effects and can grow magnetic fields larger than the ones on resonant scales. In principle, a new intriguing picture is outlined: low-energy CRs can grow fields responsible for scattering high-energy CRs. The phenomenological consequences of these findings will be explored in a forthcoming work.

\section{Conclusions}
\label{sec:conclusions}

We discuss the first confirmation of the role of NLLD as the main process regulating the growth of the resonant streaming instability in a high $\beta$ plasma. This phenomenon has been discussed for several decades now as possibly responsible for the diffusion of CRs on Galactic scales, and the self-confinement around sources. 

In the absence of damping processes, it is expected that the growth of the resonant modes continues until full isotropization of the particles is reached in the wave frame, a phenomenon resulting in advection of CRs with the waves. On the other hand, in the Galaxy the diffusion of CRs with $E\gtrsim 10$ GeV corresponds to super-Alfv\'enic drifts, which can be accommodated if the growth of the resonant instability is limited by damping. 

In the Galactic halo, the dominant damping process is most likely NLLD, and previous works have shown that a self-generated halo can form under these assumptions \cite{blasi+12,evoli+18b,chernyshov+24}.

In this letter, we scrutinize this picture with hybrid simulations of a periodic box of background gas in which a continuous hot, slowly drifting CR distribution is injected.
To mimic a fluid element close to the Galactic disk, we let the waves escape our box and study the final state of the system.
We find that saturation of the magnetic fields occurs while the drift speed is super-Alfvénic and we observe the typical gas heating that is associated with this damping process. 

However, contrary to the standard assumption, our simulations demonstrate that not all of the energy put into resonant magnetic fields is damped in the form of heat to the background gas: NLLD initiates an inverse cascade that grows fields on non-resonant, large scales, a novel effect that might have profound phenomenological implications. Although the fate of these large scale modes is unclear based on our simulations alone, it is intriguing that 
low energy CRs can generate fields that are able to resonantly scatter higher energy CRs.

It is also worth stressing that in our simulations lower energy particles appear to be harder to scatter, as small-scale modes are damped stronger in NLLD and a pitch angle barrier appears. The arising picture potentially constitutes a paradigm shift in how we understand self-generated CR transport.

The details of these phenomena and how they are realized in our Galaxy are beyond the scope of this letter and will be explored in future work.

\acknowledgements
Simulations were performed on computational resources provided by the University of Chicago Research Computing Center and on  TACC's Stampede3 through ACCESS Maximize allocation PHY240042.
We wholeheartedly thank Elena Amato, Anatoly Spitkovsky, Philipp Kempski, Matthew Kunz and Rouven Lemmerz for interesting and stimulating discussions, some of which took place at Aspen Center for Physics, which is supported by National Science Foundation grant PHY-2210452. 
This work of D.C. was partially supported by NASA through grants 80NSSC18K1218 and 80NSSC24K0173 and NSF through grants PHY-2010240, and AST-2009326.

The work of P.B. was partially supported by the European Union – NextGenerationEU RRF M4C2 1.1 under grant PRIN-MUR 2022TJW4EJ.

\end{document}